\documentclass{emulateapj}
\usepackage{epsfig, graphicx}   

\def\kms    {\ifmmode{{\rm \ts km\ts s}^{-1}}\else{\ts km\ts s$^{-1}$}\fi}
\def\msol   {\ifmmode{{\rm M}_{\odot}}\else{M$_{\odot}$}\fi}
\def\lsun   {\ifmmode{{\rm L}_{\odot}}\else{L$_{\odot}$}\fi}
\def\ts     {\thinspace} 
\def\ci   {\ifmmode{{\rm C}{\rm \small I}}\else{C\ts {\scriptsize I}}\fi}
\def\cone {\ifmmode{{\rm C}{\rm \small I}(1-0)}\else{C\ts {\scriptsize I}(1--0)}\fi}
\def\ctwo {\ifmmode{{\rm C}{\rm \small I}(2-1)}\else{C\ts {\scriptsize I}(2--1)}\fi}
\def\cii  {\ifmmode{{\rm [C}{\rm \small II}]}\else{[C\ts {\scriptsize II}]}\fi}
\def\aco  {\ifmmode{^{12}{\rm CO}(J=1\to0)}\else{$^{12}{\rm CO}(J=1\to0)$}\fi}
\def\bco  {\ifmmode{^{12}{\rm CO}(J=2\to1)}\else{$^{12}{\rm CO}(J=2\to1)$}\fi}
\def\m    {\ifmmode{\mu {\rm m}}\else{$\mu$m}\fi}
\def\cco  {\ifmmode{^{13}{\rm CO}(J=1\to0)}\else{$^{13}{\rm CO}(J=1\to0)$}\fi}
\def\dco  {\ifmmode{^{13}{\rm CO}(J=2\to1)}\else{$^{13}{\rm CO}(J=2\to1)$}\fi}
\def\eco  {\ifmmode{^{12}{\rm CO}(J=3-2)}\else{$^{12}{\rm CO}(J=3-2)$}\fi}
\def\hi   {\ifmmode{{\rm H}{\rm \small I}}\else{H\ts {\scriptsize I}}\fi}
\def\hii  {\ifmmode{{\rm H}{\rm \small II}}\else{H\ts {\scriptsize II}}\fi}
\def\ha   {\ifmmode{{\rm H}{\alpha}}\else{H${\alpha}$}\fi}
\def\hh     {\ifmmode{{\rm H}_2}\else{H$_2$}\fi}
\def\nhh     {\ifmmode{N({\rm H}_2)}\else{$N$(H$_2$)}\fi}
\def\tex {\ifmmode{{T}_{\rm ex}}\else{$T_{\rm ex}$}\fi}
\def\tmb {\ifmmode{{T}_{\rm mb}}\else{$T_{\rm mb}$}\fi}
\def\tkin {\ifmmode{{T}_{\rm kin}}\else{$T_{\rm kin}$}\fi}
\def\microns {\ifmmode{\mu{\rm m}}\else{$\mu$m}\fi}
\def\nhh   {\ifmmode{n({\rm H}_2)}\else{$n$(H$_2$)}\fi}

\slugcomment{Accepted for publication in ApJ}
\shorttitle{Constraints on the Cosmic H$_2$ Density in the HDF--N}
\shortauthors{Walter et al.}

\begin{document}

\title{A Molecular Line Scan in the  Hubble Deep Field North:\\ Constraints on the CO Luminosity Function and the Cosmic H$_2$ Density}
  
\author{
Walter F.\altaffilmark{1},
Decarli R.\altaffilmark{1}, 
Sargent M.\altaffilmark{2},
Carilli C.\altaffilmark{3},
Dickinson M.\altaffilmark{4},
Riechers D.\altaffilmark{5},
Ellis R.\altaffilmark{6},
Stark D.\altaffilmark{7},
Weiner B.\altaffilmark{7},
Aravena M.\altaffilmark{8,9}, 
Bell E.\altaffilmark{10},
Bertoldi F.\altaffilmark{11},
Cox P.\altaffilmark{12},
Da Cunha E.\altaffilmark{1},
Daddi E.\altaffilmark{4},
Downes D.\altaffilmark{12},
Lentati L.\altaffilmark{13},
Maiolino R.\altaffilmark{13},
Menten K.M.\altaffilmark{14},
Neri R.\altaffilmark{12},
Rix H.--W.\altaffilmark{1},
Weiss A.\altaffilmark{14}
}
\altaffiltext{1}{Max-Planck Institut f\"{u}r Astronomie, K\"{o}nigstuhl 17, D-69117, Heidelberg, Germany. E-mail: {\sf walter@mpia.de}}
\altaffiltext{2}{Laboratoire AIM, CEA/DSM-CNRS-Universite Paris Diderot, Irfu/Service d'Astrophysique, CEA Saclay, Orme des Merisiers, 91191 Gif-sur-Yvette cedex, France}
\altaffiltext{3}{NRAO, Pete V.\,Domenici Array Science Center, P.O.\, Box O, Socorro, NM, 87801, USA}
\altaffiltext{4}{National Optical Astronomy Observatory, 950 North Cherry Avenue, Tucson, Arizona 85719, USA}
\altaffiltext{5}{Cornell University, 220 Space Sciences Building, Ithaca, NY 14853, USA}
\altaffiltext{6}{Astronomy Department, California Institute of Technology, MC105-24, Pasadena, California 91125, USA}
\altaffiltext{7}{Steward Observatory, University of Arizona, 933 N. Cherry St., Tucson, AZ  85721, USA}
\altaffiltext{8}{European Southern Observatory, Alonso de Cordova 3107, Casilla 19001, Vitacura Santiago, Chile}
\altaffiltext{9}{N\'{u}cleo de Astronom\'{\i}a, Facultad de Ingenier\'{\i}a, Universidad Diego Portales, Av. Ej\'{e}rcito 441, Santiago, Chile}
\altaffiltext{10}{Department of Astronomy, University of Michigan, 500 Church St., Ann Arbor, MI 48109, USA}
\altaffiltext{11}{Argelander Institute for Astronomy, University of Bonn, Auf dem H\"{u}gel 71, 53121 Bonn, Germany}
\altaffiltext{12}{IRAM, 300 rue de la piscine, F-38406 Saint-Martin d'H\`eres, France}
\altaffiltext{13}{Cavendish Laboratory, University of Cambridge, 19 J J Thomson Avenue, Cambridge CB3 0HE, UK}
\altaffiltext{14}{Max-Planck-Institut f\"ur Radioastronomie, Auf dem H\"ugel 69, 53121 Bonn, Germany}

\begin{abstract} 

We present direct constraints on the CO luminosity function at high redshift and the resulting cosmic evolution of the molecular gas density, $\rho_{\rm H2}$(z), based on a blind molecular line scan in the Hubble Deep Field North (HDF--N) using the IRAM Plateau de Bure Interferometer. Our line scan of the entire 3\,mm window (79--115 GHz) covers a cosmic volume of $\sim7000$ Mpc$^3$, and redshift ranges z$<$0.45, 1.01$<$z$<$1.89 and z$>$2. We use the rich multiwavelength and spectroscopic database of the HDF--N to derive some of the best constraints on CO luminosities in high redshift galaxies to date. We combine the blind CO detections in our molecular line scan (presented in a companion paper) with stacked CO limits from galaxies with available spectroscopic redshifts (slit or mask spectroscopy from Keck and grism spectroscopy from HST) to give first blind constraints on high--z CO luminosity functions and the cosmic evolution of the H$_2$ mass density $\rho_{\rm H2}$(z) out to redshifts z$\sim$3. A comparison to empirical predictions of $\rho_{\rm H2}$(z) shows that the securely detected sources in our molecular line scan already provide significant contributions to the predicted $\rho_{\rm H2}$(z) in the redshift bins $\langle$z$\rangle\sim$1.5 and $\langle$z$\rangle\sim$2.7. Accounting for galaxies with CO luminosities that are not probed by our observations results in cosmic molecular gas densities $\rho_{\rm H2}$(z) that are higher than current predictions. We note however that the current uncertainties (in particular the luminosity limits, number of detections, as well as cosmic volume probed) are significant, a situation that is about to change with the emerging ALMA observatory.

\end{abstract}


\section{Introduction}

The last decade has seen impressive advances in our understanding of
galaxy formation and evolution based on deep field studies at various
wavelengths.  In particular, the cosmic history of star formation, and
the build up of stellar mass as a function of galaxy type and mass,
have been well quantified, starting within 1\,Gyr of the Big Bang. It
has been shown that the comoving cosmic star formation rate density
rose gradually from early epochs (at least z$\sim$6--8) to a peak
level between z$\sim$3 and~1, after which it dropped by an order of
magnitude towards the present (Hopkins \& Beacom 2006, Bouwens et al.\
2010). The build-up of stellar mass (i.e. the temporal integral)
follows this evolution (Ilbert et al. 2009; Bell et al. 2007). The
redshift range z$\sim$1--3 constitutes the `epoch of galaxy assembly',
when roughly half the stars in the Universe formed.

While progress in deep field studies has been impressive, current
knowledge of the formation of the general galaxy population is based
almost exclusively on optical, near--IR and cm--radio deep field
surveys of stars, star formation, and ionized gas. For example, Lyman Break
selected samples have revealed a major population of star--forming
galaxies at z$\sim$3 (e.g., Steidel et al.\ 2004). Likewise,
magnitude--selected samples (e.g., Le F\`evre et al.\ 2005, Lilly et
al.\ 2007) provide a census of the star-forming population based on
UV/optical flux rather than color. Radio--selected sources provide
estimates of dust--unbiased star formation rates (e.g., Cowie et al.\
2004, Dunne et al.\ 2009, Karim et al.\ 2011).

\begin{figure*} \centering
\includegraphics[width=13.0cm,angle=0]{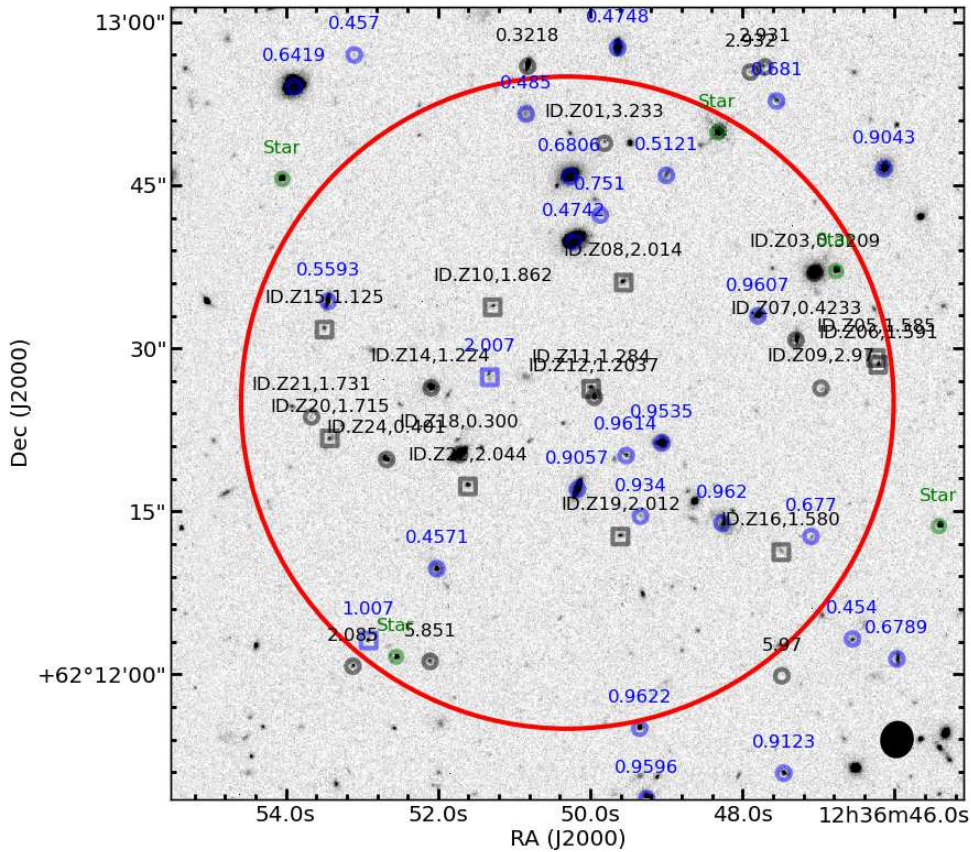} 
\caption{HST/WFC3 F160W (1.6$\mu$m) image from the CANDELS survey (Grogin et al. 2011; Koekemoer et al. 2011) of the region of the HDF--N coverd by our line scan from the CANDELS survey. The red circle shows the primary beam FWHM of our observations at the intermediate frequency of our scan (97.25 GHz). The black ellipse in the bottom-right corner shows the sythesized beam of our observations. Galaxies are labeled with their redshift. Blue colors indicate redshifts that are not covered by the frequency coverage of our 3\,mm scan (Tab.~1). Circles indicate ground--based redshifts and squares indicate slit--less (grism) redshifts (when both are available, only ground--based redshifts are shown). Green color indicates stars in the field. We show the spectroscopic completeness as a function of $H$--band magnitude in Fig.~2 and CO spectra towards all galaxies with redshift information in Fig.~3.} 
\end{figure*}

The molecular gas content is the cause of the cosmic star
formation history. However, observations of the gas content have to
date been limited to follow--up studies of galaxies that are
pre--selected from optical/NIR deep surveys (or, in the extreme cases
of quasar host galaxies and sub--millimeter galaxies, through
selection in the sub--millimeter continuum, Carilli \& Walter 2013).
In all cases the selection is based on the star formation properties
of a given galaxy.

In order to obtain an unbiased census of the molecular gas content in
high--z galaxies, there is a clear need for a blind search of
molecular gas down to mass limits characteristic of the normal star
forming galaxy population, i.e., a molecular deep field. Such a
molecular deep field has been out of reach using past instrumentation,
both in terms of sensitivity and instantaneous bandwidth. However they
are now becoming feasible, in particular given the unparalleled
sensitivity of Atacama Large (Sub--)Millimeter Array (ALMA). We here
present results based on a precursor program, using the IRAM Plateau
de Bure Interferometer (PdBI), of the Hubble Deep Field North (HDF--N,
Williams et al. 1996), that is discussed in detail in Decarli et al.\
(2013, hereafter D13). After a brief summary of the observations
(Sec.~\ref{OBS}) we discuss stacked molecular gas limits (based on
galaxies with known spectroscopic redshifts, Sec.~3.1). Together with
the `blind' CO line detections from D13 (Sec.~3.2) these give first
constraints on the redshift dependence of the CO luminosity function
in the HDF--N and their implications for the cosmic evolution of the
molecular gas content in galaxies (Sec.~\ref{OMEGA}). A short summary
and outlook is presented in Sec.~\ref{SUMMARY}. Throughout the paper
we adopt a standard cosmology with $H_0=70$ km s$^{-1}$ Mpc$^{-1}$,
$\Omega_{\rm m}=0.3$ and $\Omega_{\Lambda}=0.7$.

\section{Data}
\label{OBS}

\subsection{Complete frequency scan of the 3--mm band}
\label{SCAN}

We have observed the full 3--mm band of the PdBI
($\sim$79.7--114.8\,GHz) to approximately uniform sensitivity,
reaching an average noise of $\sim$0.3\,mJy\,beam$^{-1}$ in a
90\,km\,s$^{-1}$ channel (pointing centre: 12:36:50.300
+62:12:25.00). Observations were done in C--array configuration,
resulting in an average beam size of $\sim$\,3$"$, or $\sim$25\,kpc at
redshifts $\gtrapprox1$. At this resolution we do not expect to
spatially resolve high--redshift galaxies.  The observational details
are discussed in D13. Table 1 summarizes the redshift ranges probed by
the different CO transitions covered by our scan and Fig.~1 shows the
field covered by our observations. Table 1 also gives the cosmic
volume probed by our observations. Here we take into account that the
covered sky area, as defined by the primary beam, changes as a
function of frequency.

\begin{figure} \centering
\includegraphics[width=9.0cm,angle=0]{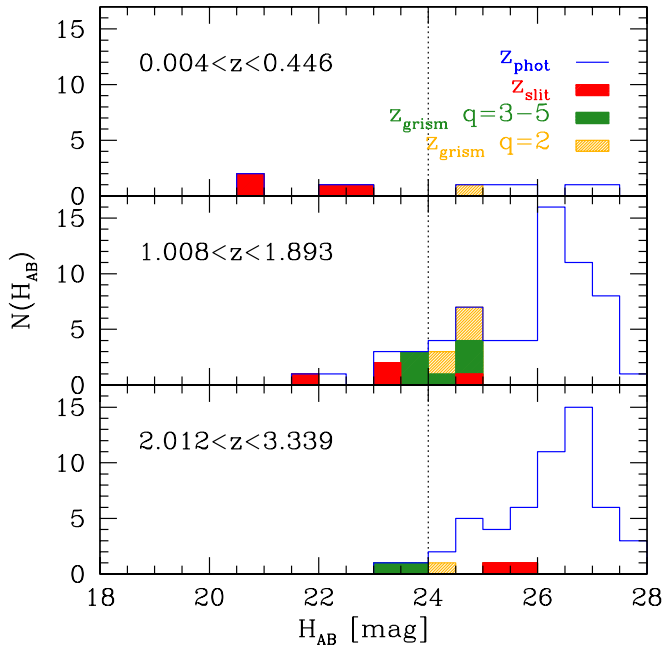} 
\caption{Histogram of number of galaxies covered in our line scan as a function of $H$--band magnitude (x--axis) and redshift bin (3 panels). The blue line shows the distribution of galaxies with photometric redshifts in each redshift bin (available for most of the galaxies in the field), whereas the colored regions indicate the availabilty of ground--based spectroscopic redshifts (red) and high--quality HST grism spectroscopy (green). Grism spectra with quality $q$=2 (yellow) are the least reliable and we do not use them for our analysis. Ground--based redshifts are preferred to HST grism ones, when both are available.} 
\end{figure}

\begin{deluxetable}{lllll}
\tabletypesize{\scriptsize}
\tablecaption{Redshift range and cosmic volume covered by molecular line scan.}
\tablewidth{0pt}
\tablehead{
\colhead{line}  & \colhead{z$_{\rm min}$} & \colhead{z$_{\rm max}$} & \colhead{$<z>$\tablenotemark{a}}  & \colhead{Volume\tablenotemark{b}}  \\
\colhead{}  &   \colhead{}  &  \colhead{}   & \colhead{}  & \colhead{Mpc$^3$}
}
\startdata
CO(1--0)  & 0.0041 & 0.446&  0.338 & 91.66 \\
CO(2--1)  &   1.01 & 1.89 &  1.52  & 1442  \\
CO(3--2)  &   2.01 & 3.34 &  2.75  & 2437  \\ 
CO(4--3)  &   3.02 & 4.78 &  3.98  & 2966  \\
CO(5--4)  &   4.02 & 6.23 &  5.21  & 3249  \\  
\enddata
\tablecomments{}
\tablenotetext{a}{Volume--averaged redshift of CO transition.}
\tablenotetext{b}{Cosmic comoving volume probed by redshift range. As sky area we use the frequency--dependent size of the PdBI primary beam (FWHM\,=\,55$"\times\frac{86}{\nu (GHz)})$.}
\end{deluxetable}

\subsection{Optical/NIR spectroscopy in the HDF--N}
\label{SPECTRA}
In our analysis we use available multiwavelength information of the
galaxies in the HDF--N, in particular (spectroscopic) redshift
estimates, to improve our sensitivity to search for CO emission.

The $H$-band selected catalogue by Dickinson et al.\ (2003), based on
deep HST/NICMOS F160W photometry, lists 220 galaxies within 30$''$
(i.e. roughly the size of the primary beam from the pointing center of
our observations. Cowie et al.\ (2004) and Barger et al.\ (2008) provide 
spectroscopic redshifts
for 15 of these. We add to this so far unpublished spectroscopic
redshifts (based on Keck spectroscopy, Dickinson et al., in
prep.) for an additional 8 galaxies up to z\,=\,4. One additional
faint galaxy is included at z\,=\,4.355 -- this redshift is based on
one line (presumably Ly--$\alpha$) and no continuum is seen in the
spectrum (Stark et al.\ 2010). All spectroscopic redshifts
based on ground observations are from the Keck telescope. In
addition to these, we add secure grism--based redshifts (based on the
detection of emission lines) and lower--quality redshifts (e.g.,
based on absorption features or on the shape of the continuum
emission) from the HST survey `A Grism H--Alpha SpecTroscopic survey'
(AGHAST, Weiner et al., in prep.).  A quality flag $q$ is assigned to all
grism-based redshifts in Tab.~2. Higher values ($q$=3--5) are
associated with grism redshifts based on emission lines. $q$=2 values
are associated with more uncertain redshifts, e.g., based on the shape
of the continuum emission. Out of our complete spectroscopic set of 47 galaxies, 27 have
a redshift that is covered by our scan (Tab.~1 and Fig.~1).

\begin{deluxetable}{cccccc}
\tabletypesize{\scriptsize}
\tablecaption{Galaxies with ground--based or HST grism--based redshifts covered by molecular line scan.}
\tablewidth{0pt}
\tablehead{
\colhead{ID}  & \colhead{RA}          & \colhead{Dec}         & \colhead{$z_{\rm spec}$} & \colhead{$z_{\rm grism}$} & \colhead{grism quality} \\
   & (J2000.0)   & (J2000.0)   &          & &     \\
\colhead{(1)} & \colhead{(2)}        & \colhead{(3)}         & \colhead{(4)}      & \colhead{(5)} & \colhead{(6)}
}
\startdata
ID.Z01 & 12:36:49.81 & +62:12:48.8 & 3.233  &         &    \\
ID.Z02 & 12:36:50.26 & +62:12:49.6 &        & [1.625] &  2 \\
ID.Z03 & 12:36:47.04 & +62:12:36.9 & 0.3209 & [0.321] &  2 \\ 
ID.Z04 & 12:36:47.61 & +62:12:37.2 &        & [0.423] &  2 \\
ID.Z05 & 12:36:46.24 & +62:12:29.1 &        & 1.585   &  3 \\
ID.Z06 & 12:36:46.22 & +62:12:28.5 &        & 1.591   &  3 \\
ID.Z07 & 12:36:47.28 & +62:12:30.7 & 0.4233 &         &    \\ 
ID.Z08 & 12:36:49.56 & +62:12:36.1 &        & 2.014   &  3 \\
ID.Z09 & 12:36:46.94 & +62:12:26.1 & 2.970  & 3       &  3 \\ 
ID.Z10 & 12:36:51.28 & +62:12:33.8 &        & 1.862   &  3 \\
ID.Z11 & 12:36:49.99 & +62:12:26.3 &        & 1.284   &  3 \\
ID.Z12 & 12:36:49.95 & +62:12:25.5 & 1.204  & 1.205   &  5 \\ 
ID.Z13 & 12:36:50.35 & +62:12:23.0 &        & [1.185] &  2 \\
ID.Z14 & 12:36:52.09 & +62:12:26.3 & 1.224  & 1.166   &  3 \\
ID.Z15 & 12:36:53.49 & +62:12:31.7 &        & 1.125   &  3 \\
ID.Z16 & 12:36:47.49 & +62:12:11.2 &        & 1.58    &  3 \\
ID.Z17 & 12:36:51.74 & +62:12:21.4 &        & [2.713] &  2 \\
ID.Z18 & 12:36:51.71 & +62:12:20.2 & 0.300  &         &    \\ 
ID.Z19 & 12:36:49.60 & +62:12:12.7 &        & 2.012   &  3 \\
ID.Z20 & 12:36:53.42 & +62:12:21.7 &        & 1.715   &  4 \\
ID.Z21 & 12:36:53.66 & +62:12:23.7 & 1.731  & 1.739   &  3 \\
ID.Z22 & 12:36:51.61 & +62:12:17.3 &        & 2.044   &  5 \\
ID.Z23 & 12:36:53.91 & +62:12:24.5 &        & [1.797] &  2 \\
ID.Z24 & 12:36:52.67 & +62:12:19.8 & 0.401  &         &    \\ 
ID.Z25 & 12:36:48.80 & +62:12:02.1 &        & [1.038] &  2 \\
ID.Z26 & 12:36:51.89 & +62:12:08.1 &        & [1.144] &  2 \\
ID.Z27 & 12:36:50.48 & +62:12:50.4 & 4.345  &         &    \\ 
\enddata
\tablecomments{Catalogue of the galaxies with ground--based or HST grism--based redshift
(from optical/NIR observations) consistent with the CO redshift coverage of our line scan (Tab.~1). (1) Line ID. (2-3) Right ascension and declination (J2000).  (4) Spectroscopic (ground--based) redshift from Cowie et al.~(2004); Reddy et al.~(2006); Barger et al.~(2008); Stark et al.~(2010). (5) Grism-based redshift from AGHAST (Weiner et al., in prep.). (6) Quality of the grism redshift (5: highest, 2: lowest; we consider only quality 3--5 in our analysis and have but the quality 2 redshifts in brackets).}
\end{deluxetable}

This final sample of 27 galaxies with redshifts covered by
our frequency scan is listed in Tab.~2. These galaxies and their
respective ID's are marked by circles (ground--based redshift) and squares
(slit--less HST grism redshifts) in Fig.~1. In Fig.~2 we show the
spectroscopic completeness in the field as a function of $H$--band
magnitude (to first order a measure of the stellar mass) for the
redshift intervals covered by our molecular line scan. From this we
conclude that we reach high spectroscopic completeness 
(i.e. $>$\,90\%) down to $H$--band magnitudes of $H_{\rm
AB}\!<\,$24\,mag for all redshift bins. This corresponds to the
following stellar masses in each redshift bin: $<$z$>$=0.338:
$\sim$5.0$\times$10$^7$\,M$_\odot$, $<$z$>$=1.52:
$\sim$3.3$\times$10$^9$\,M$_\odot$, $<$z$>$=2.75:
$\sim$7.0$\times$10$^9$\,M$_\odot$ (da Cunha et al.\ 2013).

\begin{deluxetable*}{lllllccc}
\tabletypesize{\scriptsize}
\tablecaption{Stacked CO limits based on the available spectroscopic redshift information.}
\tablewidth{0pt}
\tablehead{
\colhead{line} & \colhead{$<z>$} & \colhead{\#\tablenotemark{a}} & \colhead{S$_{\rm CO}$} &  \colhead{L$_{\rm CO}$} &  \colhead{L$'_{\rm CO}$} &  \colhead{L$'_{\rm CO(1-0)}$\tablenotemark{b}} & \colhead{density\tablenotemark{c}}\\
\colhead{} & \colhead{} & \colhead{} & \colhead{Jy\,km\,s$^{-1}$} & \colhead{10$^6$\,L$_\odot$} & \colhead{10$^9$\,K\,km\,s$^{-1}$\,pc$^2$} & \colhead{10$^9$\,K\,km\,s$^{-1}$\,pc$^2$} & \colhead{10$^{-3}$\,Mpc$^{-3}$}
}\startdata
CO(1-0) &  0.338 & 4  & $<$0.177 & $<$0.034 &  $<$0.68 & $<$0.68 & 43\\
CO(2-1) &  1.52  & 10 & $<$0.174 & $<$2.1   &  $<$5.2  & $<$5.91 & 6.9\\
CO(3-2) &  2.75  & 5  & $<$0.443 & $<$22.0  &  $<$17   & $<$34   & 2.1
\enddata
\tablecomments{All upper limits are 5\,$\sigma$. The CO(2--1) and CO(3--2) limits account for the higher uncertainties in the grism redshifts (see Sec.~3.1).}
\tablenotetext{a}{Number of galaxies in stack (from Tab.~2 and Fig.~2).}
\tablenotetext{b}{We have converted our high--J CO L$'_{\rm CO}$ luminosity limits to L$'_{\rm
CO(1-0)}$ assuming L$'_{\rm CO(2-1)}$/ L$'_{\rm CO(1-0)}$=0.84 and
L$'_{\rm CO(3-2)}$/ L$'_{\rm CO(1-0)}$=0.5 (Dannerbauer et al.\ 2009).}
\tablenotetext{c}{Volume density of sources in redshift bin using column~5 in Tab.~1.}
\end{deluxetable*}

\section{Analysis}
\label{RES}

We base our analysis on two measurements: (A) deep stacked CO limits
based on the available optical/NIR spectroscopy (Sec.~3.1) and (B) the
blind CO detections discussed in D13 (Sec.~3.2).

\subsection{Stacked CO limits based on known spectroscopic redshifts}
\label{STACK} 
\begin{figure*} \centering
\includegraphics[angle=0,width=0.7\textwidth]{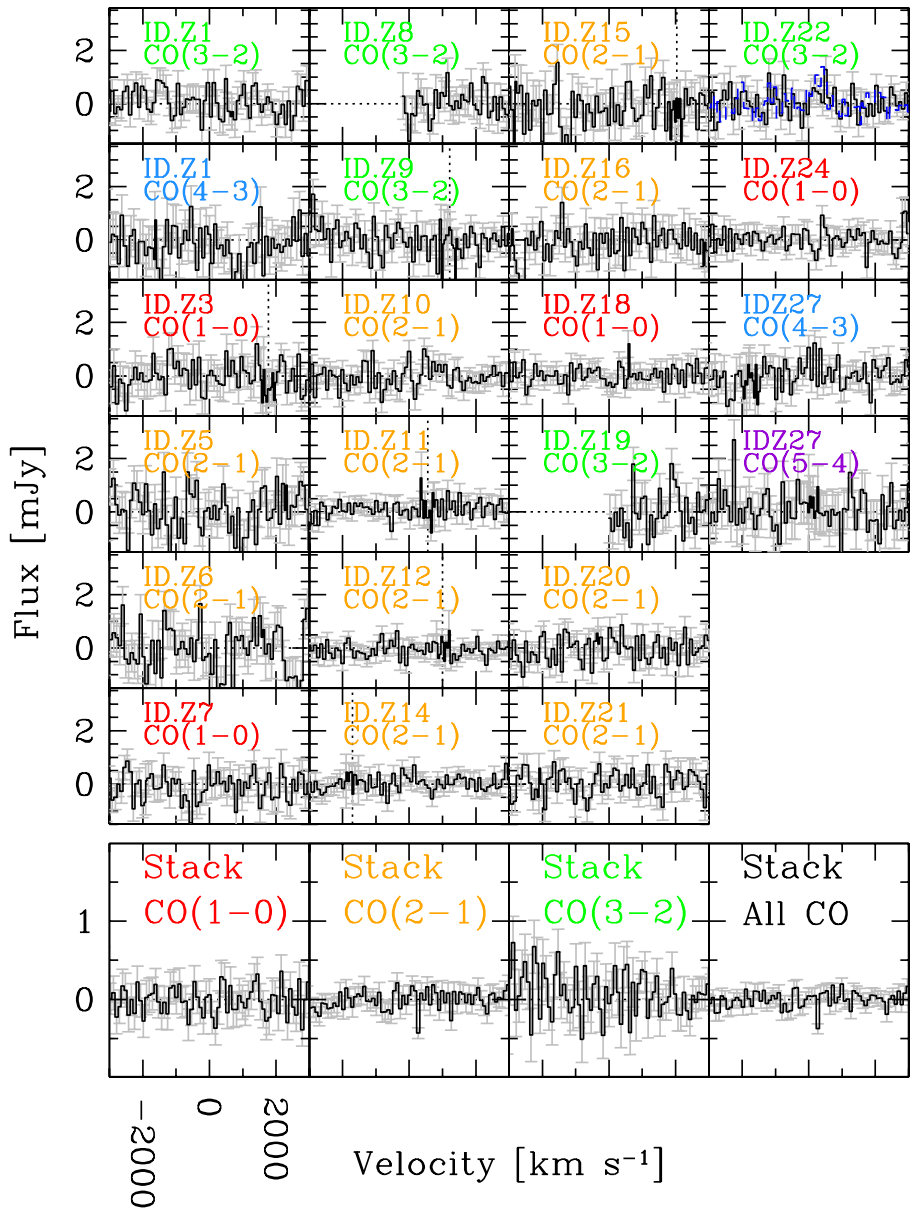}\\
\caption{Spectra of the galaxies with high--quality spectroscopic redshifts falling within the range of redshifts our scan covered for various CO transitions (see Tabs.~1 and~2). Spectra are corrected for primary beam attenuation. No galaxy is individually detected at high significance. Vertical dashed lines indicate band edges in our scan (D13). ID.Z22 is spatially coincident with our blind CO detection ID.19 (see discussion in Sec.~3.1)  and we show the spectrum of ID.19, extracted 1.5$"$ away from the optical positions, as a blue--dashed line for reference. The bottom panels show the stacked spectra for each transition and a stack of all CO emission. We note that ID.Z1 enters the latter stack twice as it has two lines in our scan. }
\label{fig_zspec}
\end{figure*}

We here use the spectroscopic redshift information
presented in Sec.~\ref{SPECTRA} to aid in our search for CO emission,
and to obtain a stacked CO limit in the galaxy samples. In
Fig.~\ref{fig_zspec} we show the spectra extracted at the pixel
of the nominal positions of the galaxies with spectroscopic redshifts
(Tab.~2), shifted to their respective redshifts; here we exclude
sources that only have low--quality (quality 2) grism redshifts. 
All spectra have been corrected for primary beam attenuation, leading
to different noise properties in the spectra. None of the spectra
show convincing CO emission at the expected redshift. In our blind
search for CO (D13) we report a candidate CO line emission for one of
the sources shown in Fig.~\ref{fig_zspec}, ID.Z22, which is spatially
consistent with the CO line candidate ID.19 and where the grism
redshift matches the CO redshift perfectly (see detailed discussion in
D13)\footnote{For this source, we also show the spectrum that
corresponds to the CO candidate ID.19 that is offset by 1.5$"$ from
the optical/NIR counterpart for the optical galaxy ID.Z22 in
Fig.~\ref{fig_zspec}. See detailed discussion of CO candidate ID.19 in
D13.}.  We also note that one galaxy, ID.Z27 at z=4.345 (Stark et al.\ 2010),
shows a tentative CO(4--3) line but we treat this as an upper limit in
our analysis.

To stack the spectra, we first need to consider the accuracy of the
available optical/NIR redshifts: The typical uncertainties of
Keck spectroscopic redshifts $z\leq1.6$ are of order few tens of \kms{}
(Newman et al.\ 2012), and we consider these uncertainties neglibile
for our stacking, given the expected line widths of
$\sim$\,300\,km\,s$^{-1}$ (e.g., Carilli \& Walter 2013). At higher
redshifts, the uncertainties are higher (a few hundred \kms) due to
various observational and astrophysical biases e.g., lack of bright
nebular lines, such as [O\,III] or [O\,II], or rest frame UV features
showing systematic offsets (e.g. Steidel et al.\ 2010). The average
uncertainty in the grism redshifts is $\delta z/(1+z)\approx0.0016$
(Weiner et al., in prep.). These higher uncertainties are
related to poorer spectral resolution, confusion between spatial and
spectral structure in slitless observations, and the intrinsic
weakness of the lines. 

For our stack we weight--average the primary--beam corrected
spectra after realignment and rebinning (bottom panels of
Fig.~\ref{fig_zspec}). Given the uncertainties in grism
redshifts, we compute the stacked flux as the integral (and its
uncertainties) of the stacked spectra over 1000\,\kms (i.e.,
sufficient to encompass any CO emission within the typical redshift
uncertainties). A tighter velocity range of 300\,\kms\ was assumed for
the lowest redshift bin, where all galaxies have a more accurate
ground--based redshift. The final stacked upper limits for the CO fluxes
and resulting luminosities are given in Tab.~3.

\subsection{Blind CO detections from molecular line scan}
\label{BLIND}

\begin{deluxetable*}{lllccl}
\tabletypesize{\scriptsize}
\tablecaption{Number of blind CO detections and limiting luminosities in scan}
\tablewidth{0pt}
\tablehead{
\colhead{line} & \colhead{$<z>$}  & \colhead{\#\tablenotemark{a}} & \colhead{L'$_{\rm CO}^{3 \sigma}$} & \colhead{L'$_{\rm CO(1-0)}^{3\sigma}$} & \colhead{density\tablenotemark{c}} \\
\colhead{} & \colhead{}  & \colhead{} & \colhead{10$^9$\,K\,km\,s$^{-1}$\,pc$^2$} & \colhead{10$^9$\,K\,km\,s$^{-1}$\,pc$^2$}\tablenotemark{b} & \colhead{10$^{-3}$\,Mpc$^{-3}$}
}
\startdata
CO(1--0)  &  0.338 & 0    & 1.0  & 1.0  & $<$10.9     \\
CO(2--1)  &  1.52  & 1--3 & 5.2  & 6.2  & 0.69--2.10  \\
CO(3--2)  &  2.75  & 1--8 & 6.6  & 13.2 & 0.41--3.28  \\
\enddata
\tablenotetext{a}{Number of blind detections in the molecular line scan above our sensitivity limit (next column) as derived in D13 (see Sec 3.2).}
\tablenotetext{b}{See Tab.~3 caption for details on conversion from L$'_{\rm CO}$ to L$'_{\rm CO(1-0)}$.}
\tablenotetext{c}{Volume density of sources in redshift bin using column 2 in Tab.~1. In case of no detection (first redshift bin) we assume an upper limit of one source.}
\end{deluxetable*}

In D13 we present a blind search for CO emission in the molecular line
scan above a luminosity limit of
$\sim$6$\times$10$^9$\,K\,km\,s$^{-1}$\,pc$^2$ (to first order
irrespective of CO transition, see D13). 17 potential candidate lines
are discussed in D13 and here we concentrate on the ones with quality
flag `high--quality' and `secure', which leaves 13 sources. Two line
candidates (ID.08 and ID.17) belong to the highly obscured galaxy
HDF\,850.1 at z=5.183 (Walter et al.\ 2012). This source has
previously been selected as a sub--millimeter galaxy (and our field
centre was chosen to include it) so we do not discuss it further
here. There are two addtional galaxies in the `secure' list for which
we are certain of their redshifts: ID.03 at z=1.7844 (redshift derived
from three CO lines) and ID.19 at z=2.0474 (ID.Z22 in Tab.~2; coincident CO and
grism redshift); in both cases the spectral energy distributions
(SEDs) based on the available multi--wavelength photometry are in
excellent agreement with the derived redshifts (D13).

Based on the available multi--wavelength information, D13 assigned
each of the remaining line candidates a tentative redshift. We stress
that we expect some of the line candidates to be spurious
\footnote{In D13 we have derived the number of likely spurious
detections using simulated data cubes. We find 2 sources classified as
'high--quality/secure', i.e. with spectroscopic S/N$>$3.5, that 
are false detections. If we assign one to each
redshift bin this leads to a spurious fraction of 1/3 and 1/8 for the
two highest redshift bins.}, and consequently treat the number of
candidate detections in each redshift bin as an upper limit. Dedicated
follow--up observations in other CO transitions are needed to confirm
the reality and redshifts of our candidate lines. We record the number
of blind line detections and limiting magnitudes in each redshift bin
in Tab.~4.

\section{Implications}
\label{OMEGA}

\begin{figure*} \centering
\includegraphics[angle=-90,width=17.0cm]{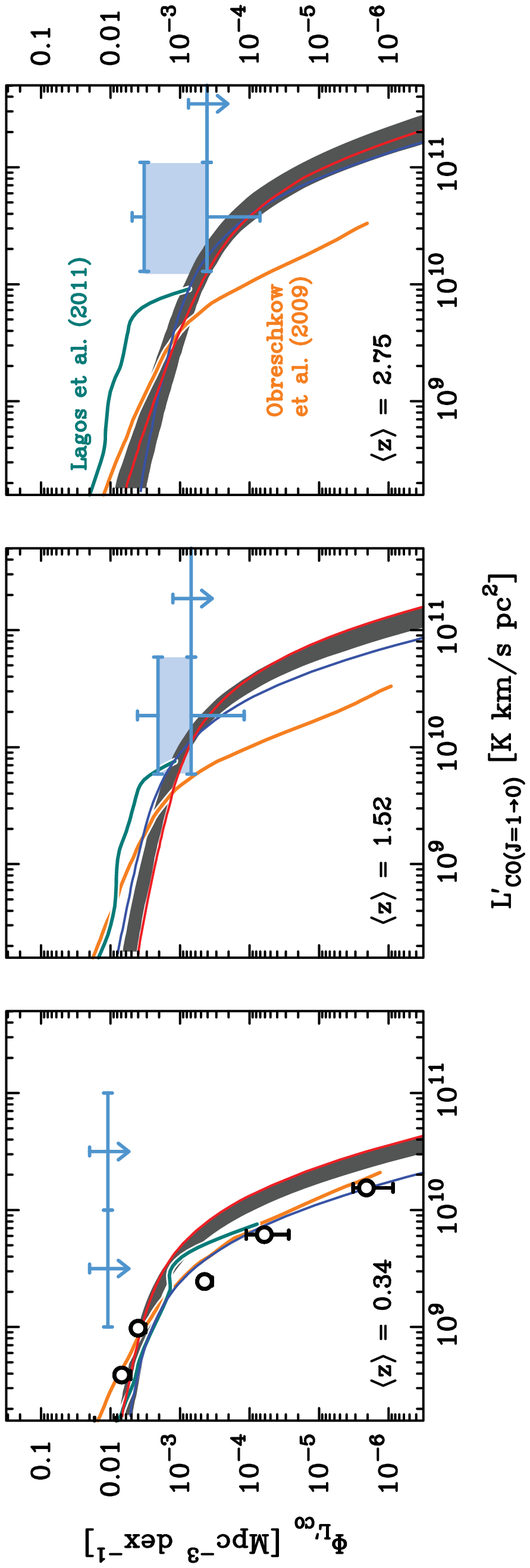}
\caption{Comparison of our blind CO measurements with 
empirically derived CO luminosity functions from the literature. The
grey shading shows the predictions by Sargent et
al. (2013a) for the average (volume--weighted) redshift (Tab.~1) where
the width indicates the 68\% confidence region. As each redshift bin
covers a significant range in redshifts we also show the median
luminosity function for the lowest and highest redshift in the
respective redshift bin (red and blue curve, here the 68\% confidence
region is not shown).  Also shown are models for the evolution of the
CO luminosity function based on semi--analytical cosmological models
plus `recipes' to relate gas mass to CO luminosity (Lagos et al.\ 2011
and Obreschkow et al.\ 2009a,b).  In the left panel the observational
constraints on the local (z\,=\,0) CO luminosity functions reported in
Kere{\v s} et al.\ (2003) are also shown as data points (small
circles). The blue--shaded regions shows the constraints derived from
our blind detections (Tab.~4, Secs.~3.2 and~4.1), including
appropriate error bars (following Gehrels 1986).}
\end{figure*}

\begin{figure*} \centering
\includegraphics[width=15.0cm,angle=0]{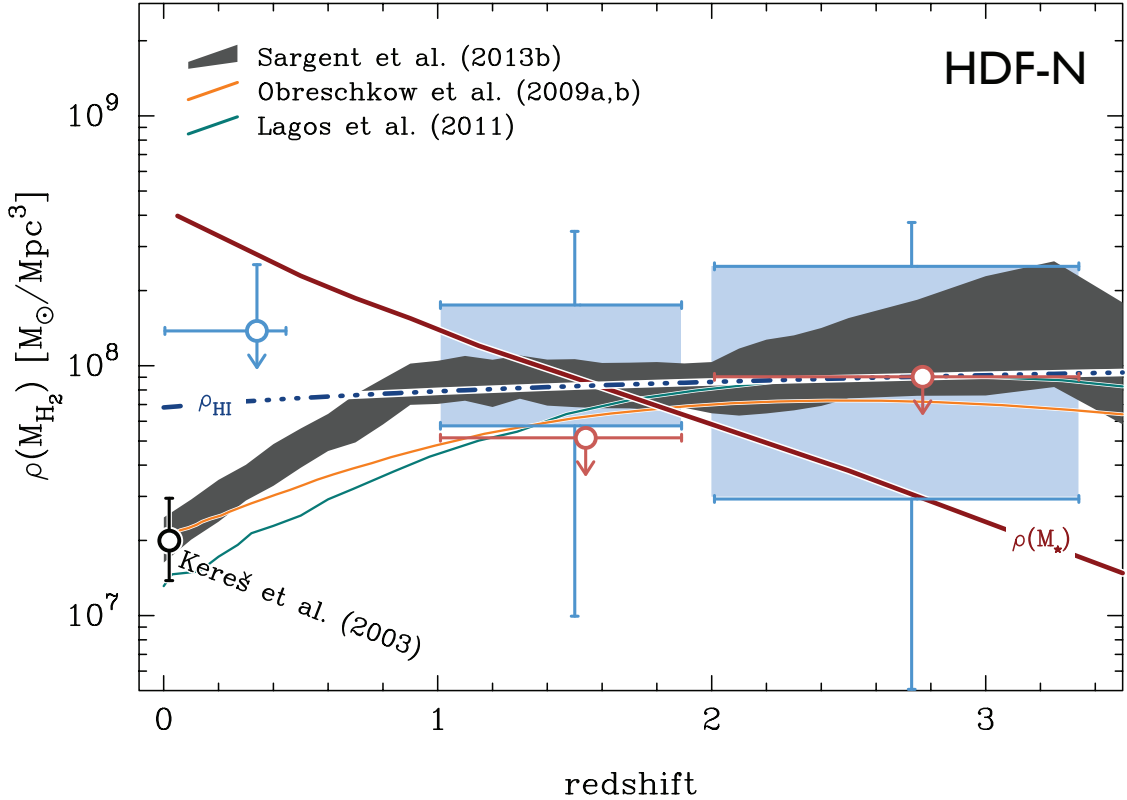} \caption{The
evolution of the cosmic H$_2$ mass density, $\rho_{\rm H2}$(z) based
on predictions from semi--analytical comological models (Obreschkow et
al.\ 2009a, 2009b, Lagos et al.\ 2011) as well as the empirical
predictions by Sargent et al.\ (2013b). The latter shows the evolution
inferred from the integration of the indirectly inferred molecular gas
mass functions underlying the CO luminosity distributions of
Fig.~4. The blue--shaded area shows only the contribution of the blind
detections (blue--shaded regions in Fig.~4) to $\rho$(M$_{\rm H2}$), {\em
not} corrected/extrapolated for a population of undetected sources at
lower or higher L$'_{\rm CO}$. The red upper limit indicates the limit
to the gas density contribution strictly for the specific galaxy
populations selected via optical spectroscopic
redshifts (Sec.~3.1). Our limit in the lowest redshift bin is not very
constraining given the small volume probed. For a comparison,
the evolution of the cosmic neutral gas mass density ($\rho_{\rm HI}$) 
and of the stellar mass density ($\rho_{\rm *}$) are also plotted.} \end{figure*}

Our molecular line scan in the HDF--N constitutes the first systematic
blind search for CO emission down to a mass limit that is
characteristic of galaxies that lie on the relatively tight `main 
sequence' SFR--$M_{\rm *}$ relation (Daddi et al.~2007). Its cosmic 
volume is well defined and characterized
through the ancillary multi--wavelength observations. In the following
we discuss our constraints on the CO luminosity functions and the
cosmic evolution of the cosmic molecular gas density $\rho_{\rm H2}$(z) in
the HDF--N.

\subsection{Constraints on CO luminosity function}

We now constrain the CO luminosity function at different redshifts
based on our blind CO detections (Tab.~4, Sec.~3.2). We compare our
measurements to empirical predictions of the CO luminosity
function. In Fig.~4 we show the CO(1--0) luminosity function in the
three redshift bins covered by our line scan, as predicted by Sargent 
et al.\ (2013b) based on (1) the evolution of the stellar
mass function of star--forming galaxies, (2) the redshift evolution of
the specific SFR of main--sequence galaxies, (3) the distribution of
main--sequence and star--bursting galaxies in the
SFR--$M_{\star}$--plane (Sargent et al. 2012), (4) distinct
prescriptions of the star formation efficiency of main--sequence and
star--bursting galaxies, and (5) a metallicity-dependent conversion
factor $\alpha_{\rm CO}$.

In the same plot (Fig.~4) we also show the predictions based on
semi--analytical cosmological models by Lagos et al.\ (2011) and
Obreschkow et al. (2009), both interpolated to our relevant median
$<$z$>$ based on their predictions at lower/higher redshift.  The
luminosities plotted here are computed for the CO(1--0) line (i.e. L$'_{\rm CO(1-0}$)
and we have converted our higher--J CO luminosities and upper limits to
L$'_{\rm CO(1-0)}$ (see Tab.~3 caption). The CO luminosity
function has to date only been directly measured at $z$\,=\,0 (see
data points by Kere{\v s} et al.\ 2003 in the left panel of Fig.~4).

In order to compare our measurements to these empirically predicted
luminosity functions, we need to normalize our number densities in a
consistent way. In the literature, the volume densities are typically
given in units of sources per Mpc$^3$ and dex in luminosity (L$'_{\rm
CO(1-0)}$). For our blind detections we define luminosity bins to
range from our 3$\sigma$ limiting luminosity (Tab.~4) to a luminosity
that is a factor of 10 higher (i.e., 1~dex). We then count the number
of blind detections in this luminosity bin. We have at least two
secure CO detections (one in the z$\sim$1.52 redshift bin, one in the 
bin with $<$z$>$=2.75), and
potentially up to 11 (3 at $<$z$>$=1.52, 8 at $<$z$>$=2.75), if we
include all the `high--quality' line candidates in D13 (see Sec.~3.2).
The lower limits in the plots are thus from our secure detections. The
upper limits represent the case where all line candidates are in fact
real. In the limit of low number statistics we adopt the Poisson
errors following Gehrels (1986, their tables~1 and~2). We plot the
allowed parameter space of our observations as blue--shaded regions in
Fig.~4. Our blind detections thus probe the `knee' of the predicted CO
luminosity functions.

\subsection{Constraints on $\rho_{\rm H2}$(z)}

We now proceed to convert our constraints on the CO luminosity
function to constraints on the cosmic volume density of the molecular
gas mass $\rho_{\rm H2}$(z). This is shown in Fig.~5 where we compare our
observations to the same models and empirical predictions as discussed
in Sec.~4.1 and Fig.~4. For a comparison we also show the evolution
of $\rho$(M$_{\rm HI}$) based on Bauermeister et al.\ (2010) and
$\rho$(M$_\star$) based on the compilations in Marchesini et
al. (2009) and Fontana et al. (2006).

The blue--shaded area in Fig.~5 (with appropriate error bars)
indicates the contribution of our blind CO detections to
$\rho_{\rm H2}$(z) from Fig.~4. To translate the observed CO luminosities
to H$_2$ masses we have asssumed a galactic H$_2$--to--CO conversion
factor and BzK excitation (see discussion in D13). In Fig.~5 we do not
attempt to correct for sources not detected in our scan at both lower
and higher L$'_{\rm CO}$ luminosities, given the unknown shape of the
luminosity function. We note though that if we simply scaled up
the predicted luminosity functions by Sargent et al. (2013b) so that
they are consistent with our observational constraints in Fig.~4, then
the values for $\rho_{\rm H2}$(z) would lie above the empirical
predictions shown in Fig.~5. This however would imply an overestimate 
of the number--density of the known population of galaxies.

In Fig.~5 we also show the contribution of the galaxies for which we
obtained a stacked upper limit (Sec.~3.1). Like in the case of the
blind detections, we do not attempt to correct this measurement for
undetected sources at lower (or higher) CO luminosities. The upper
limit shown in red color thus simply represents the contribution of
the galaxies for which spectroscopic information is available as
discussed in Sec.~3.1 and Fig.~2 and 3.

We conclude that the contribution of {\em just} our blind detections
to the cosmic molecular gas density $\rho_{\rm H2}$(z) already
consitute a major contribution to current empirical predictions and
models.  As a consequence, accounting for the contribution of yet
undetected sources (at lower or higher CO luminosities) would give
values that lie above the predictions. 

\section{Summary and Outlook}
\label{SUMMARY}

Our molecular line scan of the HDF--N (D13) allows us to place first
direct `blind' limits on the molecular gas density in `normal'
galaxies at high redshift.  We have used the rich
multiwavelength and spectroscopic database to derive some of the best
constraints on CO luminosities in high redshift galaxies to date. We
combine our blind CO detections (D13) with stacked CO limits based on
galaxies with available spectroscopic redshifts in the HDF--N to give
first constraints on the CO luminosity functions and the cosmic
evolution of the H$_2$ mass density $\rho_{\rm H2}$(z) out to
redshifts z$\sim$\,3 in the HDF--N.

The securely detected sources in our molecular line scan provide
significant contributions to the predicted $\rho_{\rm H2}$(z) in the
redshift bins $<$z$>\sim$1.5 and $<$z$>\sim$2.7. If we were to
extrapolate their contribution towards undetected sources at lower and
higher CO luminosities we would get $\rho$(M$_{\rm H2}$) values that would
exceed predictions (the precise number will depend on the assumed
shape of the luminosity function).  This in turn would imply that
current predictions, e.g., those based on the galaxies' star formation
activity and `inverting' the star formation law, would not account for
all the molecular gas that is actually present in galaxies. We note
however that the current uncertainties in our precursor study (in
particular the current luminosity limits, number of detections, as
well as cosmic volume probed) are significant, and that current models
can thus not be ruled out given the available data. In addition, the
effects of cosmic variance can not be evaluated by just looking at one
field.  The emerging capabilities of the Atacama Large
(Sub--)Millimeter Array (ALMA) with its order--of--magnitude increase
in sensitivity will enable similar molecular deep field studies to
much deeper levels and larger areas (e.g. Da Cunha et al.\ 2012,
Carilli \& Walter 2013). One caveat is that CO may break down as a
reliable tracer for H$_2$ mass in extreme environments, in particular
at low metallicities (e.g. Bolatto et al., 2013, Genzel et al.\ 2012,
Schruba et al.\ 2011). The sensitivities of our current observations
are such that we  are only sensitive to gas in galaxies as massive 
and luminous as L$^*$,
i.e. environments in which the metallicities are not expected to be
below solar given the fundamental metallicity relation (Mannucci et
al. 2011). Future, much more
sensitive, observations of molecular deep fields with ALMA that
include measurements of the dust continuum (and thus an independent
measure of the mass of the interstellar medium) are essential to (a)
further constrain possible metallicity dependences of the
CO--to--H$_2$ conversion factor for a large sample of
well--characterized high redshift galaxies and (b) further directly
constrain the cosmic evolution of the molecular gas reservoir in
galaxies.

\acknowledgements

We thank the referee for a very helpful and constructive report. FW,
DR and EdC acknowledge the Aspen Center for Physics where parts of
this manuscript were written. This paper is based on observations with
the IRAM Plateau de Bure Interferometer (PdBI). IRAM is supported by
INSU/CNRS (France), MPG (Germany) and IGN (Spain). Support for RD was
provided by the DFG priority program 1573 `The physics of the
interstellar medium'.

\label{lastpage}

\begin{thebibliography}{99}

\bibitem[\protect\citeauthoryear{Barger et al.}{2008}]{2008ApJ...689..687B} Barger A.J., Cowie L.L., Wang W.-H., 2008, \apj, 689, 687

\bibitem[\protect\citeauthoryear{Bauermeister et al.}{2010}]{2010ApJ...717..323B} Bauermeister, A., Blitz, L., \& Ma, C.-P.\ 2010, \apj, 717, 323 

\bibitem[\protect\citeauthoryear{Bell et al.}{2007}]{2007ApJ...663..834B} Bell, E.~F., Zheng, X.~Z., Papovich, C., et al.\ 2007, \apj, 663, 834 

\bibitem[\protect\citeauthoryear{Bolatto et al.}{2013}]{2013arXiv1301.3498B} Bolatto, A.~D., Wolfire, M., \& Leroy, A.~K.\ 2013, arXiv:1301.3498 

\bibitem[\protect\citeauthoryear{Bouwens et al.}{2010}]{2010ApJ...709L.133B} Bouwens, R.~J., Illingworth, G.~D., Oesch, P.~A., et al.\ 2010, \apjl, 709, L133 


\bibitem[\protect\citeauthoryear{Carilli \& Walter}{2013})]{2013arXiv1301.0371C} Carilli, C., \& Walter, F.\ 2013, arXiv:1301.0371 

\bibitem[Cowie et al.(2004)]{2004ApJ...603L..69C} Cowie, L.~L., Barger, A.~J., Fomalont, E.~B., \& Capak, P.\ 2004, \apjl, 603, L69 

\bibitem[\protect\citeauthoryear{da Cunha et al.}{2013}]{2013ApJ...765....9D} da Cunha, E., Walter, F., Decarli, R., et al.\ 2013, \apj, 765, 9 

\bibitem[\protect\citeauthoryear{Daddi et al.}{2007}]{2007ApJ...670..156D} Daddi E., et al.\ 2007, \apj, 670, 156

\bibitem[\protect\citeauthoryear{Dannerbauer et al.}{2009}]{2009ApJ...698L.178D} Dannerbauer, H., Daddi, E., Riechers, D.~A., et al.\ 2009, \apjl, 698, L178 

\bibitem[\protect\citeauthoryear{Decarli et al.}{2013}]{decarli13}Decarli, R., Walter, F., et al., ApJ, subm.\ 2013 (D13)\\

\bibitem[\protect\citeauthoryear{Dickinson et al.}{2003}]{2003ApJ...587...25D} Dickinson, M., 
Papovich, C., Ferguson, H.~C., \& Budav{\'a}ri, T.\ 2003, \apj, 587, 25 

\bibitem[Dunne et al.(2009)]{2009MNRAS.394....3D} Dunne, L., Ivison, R.~J., Maddox, S., et al.\ 2009, \mnras, 394, 3 

\bibitem[\protect\citeauthoryear{Fontana et al.}{2006}]{2006A&A...459..745F} Fontana, A., Salimbeni, S., Grazian, A., et al.\ 2006, \aap, 459, 745 

\bibitem[\protect\citeauthoryear{Gehrels}{1986}]{1986ApJ...303..336G} Gehrels, N.\ 1986, \apj, 303, 
336 

\bibitem[\protect\citeauthoryear{Ilbert et al.}{2009}]{2009ApJ...690.1236I} Ilbert, O., Capak, P., Salvato, M., et al.\ 2009, \apj, 690, 1236 

\bibitem[\protect\citeauthoryear{Genzel et al.}{2012}]{2012ApJ...746...69G} Genzel, R., Tacconi, L.~J., Combes, F., et al.\ 2012, \apj, 746, 69 

\bibitem[\protect\citeauthoryear{Grogin et al.}{2011}]{2011ApJS..197...35G} Grogin, N.~A., Kocevski, D.~D., Faber, S.~M., et al.\ 2011, \apjs, 197, 35 

\bibitem[\protect\citeauthoryear{Hopkins \& Beacom}{2006}]{2006ApJ...651..142H} Hopkins, A.~M., \& Beacom, J.~F.\ 2006, \apj, 651, 142 

\bibitem[\protect\citeauthoryear{Karim et al.}{2011}]{2011ApJ...730...61K} Karim, A., Schinnerer, E., Mart{\'{\i}}nez-Sansigre, A., et al.\ 2011, \apj, 730, 61 

\bibitem[\protect\citeauthoryear{Keres et al.}{2003}]{2003ApJ...582..659K} Keres, D., Yun, M.~S., \& Young, J.~S.\ 2003, \apj, 582, 659 

\bibitem[\protect\citeauthoryear{Koekemoer et al.}{2011}]{2011ApJS..197...36K} Koekemoer, A.~M., Faber, S.~M., Ferguson, H.~C., et al.\ 2011, \apjs, 197, 36 

\bibitem[\protect\citeauthoryear{Lagos et al.}{2011}]{2011MNRAS.418.1649L} Lagos, C.~D.~P., Baugh, C.~M., Lacey, C.~G., et al.\ 2011, \mnras, 418, 1649 

\bibitem[\protect\citeauthoryear{Le F{\`e}vre et al.}{2005}]{2005A&A...439..845L} Le F{\`e}vre, O., Vettolani, G., Garilli, B., et al.\ 2005, \aap, 439, 845 

\bibitem[\protect\citeauthoryear{Lilly et al.}{2007}]{2007ApJS..172...70L} Lilly, S.~J., Le F{\`e}vre, O., Renzini, A., et al.\ 2007, \apjs, 172, 70 

\bibitem[\protect\citeauthoryear{Marchesini et al.}{2009}]{2009ApJ...701.1765M} Marchesini, D., van Dokkum, P.~G., F{\"o}rster Schreiber, N.~M., et al.\ 2009, \apj, 701, 1765 

\bibitem[\protect\citeauthoryear{Mannucci et al.}{2010}]{2010MNRAS.408.2115M} Mannucci, F., Cresci, G., Maiolino, R., Marconi, A., \& Gnerucci, A.\ 2010, \mnras, 408, 2115 

\bibitem[\protect\citeauthoryear{Newman et al.}{2012}]{2012arXiv1203.3192N} Newman, J.~A., Cooper, M.~C., Davis, M., et al.\ 2012, arXiv:1203.3192 

\bibitem[\protect\citeauthoryear{Obreschkow et al.}{2009}]{2009ApJ...702.1321O} Obreschkow, D., Heywood, I., Kl{\"o}ckner, H.-R., \& Rawlings, S.\ 2009a, \apj, 702, 1321 

\bibitem[\protect\citeauthoryear{Obreschkow et al.}{2009}]{2009ApJ...698.1467O} Obreschkow, D., Croton, D., De Lucia, G., Khochfar, S., \& Rawlings, S.\ 2009b, \apj, 698, 1467 

\bibitem[\protect\citeauthoryear{Sargent et al.}{2013a}]{2013arXiv1303.4392S} Sargent, M.~T., Daddi, E., B{\'e}thermin, M., et al.\ 2013, arXiv:1303.4392 

\bibitem[\protect\citeauthoryear{Sargent et al.}{2013b}]{sargent13}Sargent et al. 2013b, M.~T., et al., in prep.

\bibitem[\protect\citeauthoryear{Sargent et al.}{2012}]{2012ApJ...747L..31S} Sargent, M.~T., B{\'e}thermin, M., Daddi, E., \& Elbaz, D.\ 2012, \apjl, 747, L31 

\bibitem[\protect\citeauthoryear{Schruba et al.}{2012}]{2012AJ....143..138S} Schruba, A., Leroy, 
A.~K., Walter, F., et al.\ 2012, \aj, 143, 138 

\bibitem[\protect\citeauthoryear{Stark et al.}{2010}]{2010MNRAS.408.1628S} Stark, D.~P., Ellis, R.~S., Chiu, K., Ouchi, M., \& Bunker, A.\ 2010, \mnras, 408, 1628 

\bibitem[\protect\citeauthoryear{Steidel et al.}{2004}]{2004ApJ...604..534S} Steidel, C.~C., Shapley, A.~E., Pettini, M., et al.\ 2004, \apj, 604, 534 

\bibitem[\protect\citeauthoryear{Steidel et al.}{2010}]{2010ApJ...717..289S} Steidel, C.~C., Erb, D.~K., Shapley, A.~E., et al.\ 2010, \apj, 717, 289 

\bibitem[\protect\citeauthoryear{Walter et al.}{2012}]{2012Natur.486..233W} Walter, F., Decarli, R., Carilli, C., et al.\ 2012, \nat, 486, 233 

\bibitem[\protect\citeauthoryear{Williams et al.}{1996}]{1996AJ....112.1335W} Williams, R.~E., Blacker, B., Dickinson, M., et al.\ 1996, \aj, 112, 1335 

\end{thebibliography}
\end{document}